\documentclass[prd,twocolumn,preprintnumbers,nofootinbib]{revtex4}
\usepackage{graphicx}

\usepackage[plainpages=false, colorlinks=true, anchorcolor=blue, linkcolor=blue, citecolor=blue, bookmarks=false]{hyperref}
\usepackage{color}
\newcommand{\rthis}[1]{\textcolor{black}{#1}}
\usepackage{float}
\usepackage{amsfonts,amsmath,amssymb}
\usepackage{natbib}
\usepackage{array}
\usepackage{graphicx}
\usepackage{pdfpages}
\usepackage{booktabs}
\usepackage{hhline}
\usepackage[font=footnotesize]{caption}
\usepackage{multirow}
\usepackage{url}
\begin{document}
\newcolumntype{P}[1]{>{\centering\arraybackslash}p{#1}}
\pdfoutput=1
\newcommand{\jcap}{JCAP}
\newcommand{\araa}{Annual Review of Astron. and Astrophys.}
\newcommand{\apss}{Astrophysics and Space Sciences}
\newcommand{\aj}{Astron. J. }
\newcommand{\mnras}{MNRAS}
\newcommand{\physrep}{Physics Reports}
\newcommand{\apjl}{Astrophys. J. Lett.}
\newcommand{\apjs}{Astrophys. J. Suppl. Ser.}
\newcommand{\aap}{Astron. \& Astrophys.}
\newcommand{\pasa}{PASA}
\renewcommand{\arraystretch}{2.5}
\title{Constraint on Lorentz Invariance Violation for spectral lag transition in  GRB 160625B using profile likelihood}
\author{Shantanu \surname{Desai}$^1$ }
\altaffiliation{E-mail: shntn05@gmail.com}
\author{Shalini \surname{Ganguly}$^2$} \altaffiliation{E-mail: gangulyshalini1@gmail.com}
\begin{abstract}
We reanalyze the spectral lag data  for  GRB 160625B   using frequentist inference in order to constrain the energy scale ($E_{QG}$) of Lorentz Invariance Violation (LIV). For this purpose, we use profile likelihood to deal with the astrophysical nuisance parameters.  This is in contrast to Bayesian inference implemented in previous works, where marginalization was carried out over the nuisance parameters.  We show that with profile likelihood,  we do not find a global minimum for $\chi^2$ as a function of  $E_{QG}$ below the Planck scale for both linear and quadratic models of LIV,  whereas   bounded credible intervals were previously obtained using Bayesian inference. Therefore, we can set one-sided  lower limits  in a straightforward manner. We find that  $E_{QG} \geq 2.55 \times 10^{16}$ GeV and  $E_{QG} \geq 1.85 \times 10^7$ GeV at \rthis{95\% c.l.,} for linear and quadratic LIV, respectively. Therefore, this is the first proof-of-principles application of profile likelihood method to the analysis of GRB spectral lag data to constrain LIV.
\end{abstract}

\affiliation{$^{1}$Department  of Physics, IIT Hyderabad,  Kandi, Telangana-502284, India}
\affiliation{$^{2}$Department of Physics, St. Mary's College of Maryland,  St. Mary's City, MD, 20686, USA}
\maketitle
\section{Introduction}
The spectral lags of Gamma-ray Bursts (GRBs) have been widely used~\cite{Desairev,WuGRBreview,WeiWu2} as a probe of Lorentz invariance Violation (LIV) ever since this was first proposed more than two decades ago~\cite{AmelinoCamelia98}. The spectral lag is defined as the time difference  between the arrival of high energy and low energy photons, and is positive if the high energy photons precede the low energy ones. In case of LIV caused by an energy-dependent slowing down of speed of light, one expects a turnover in the spectral lag data at high energies.

Among the plethora of searches for LIV using GRBs, the first work which found a turnover in the spectral lag data was by \citet{Wei} (W17, hereafter).  This analysis found evidence for  a transition from positive to negative time lag in the spectral lag data for GRB 160625B. By modeling the time lag as sum of intrinsic astrophysical  time-lag  and an energy-dependent speed of light,  which kicks in at high energies, they argued that this observation constitutes a robust evidence for a turnover in the spectral lag data, which could be caused by LIV. Statistical significance of this turnover was then calculated using frequentist, information theory and Bayesian model selection techniques~\cite{Ganguly,Gunapati}. Using Bayesian inference, lower limits on the quantum gravity energy scale was set at $0.5 \times 10^{16}$ GeV and $1.4 \times 10^7$ GeV for linear and quadratic LIV, respectively~\cite{Wei}. These limits were obtained by marginalizing over the astrophysical nuisance parameters. All other analyses searching for LIV using GRB spectral lags have always used Bayesian inference. These include some of our own past works~\cite{Agrawal_2021,Desai23,Pasumarti23}.

In this work we redo the analysis in ~\cite{Wei} using frequentist inference, where we deal with the astrophysical nuisance parameters using profile likelihood. While the profile likelihood is a ``bread and butter'' tool in experimental high energy Physics~\cite{PDG}, until recently it has seldom been used in Astrophysics, where Bayesian inference is commonly used. Recently, however there has been a renaissance in the use of profile likelihood in the field of Cosmology (see ~\cite{Herold,Campeti,Colgain24,Karwal24,Herold24} for an incomplete list). In particular case, it was shown that one reaches opposite conclusions for the fraction of Early Dark energy using Profile Likelihood as compared to Bayesian inference~\cite{Herold}.

The outline of this manuscript is as follows. We review the basic data analysis done in W17 to search for LIV  in Section~\ref{sec:data}. We compare and contrast Bayesian and frequentist parameter estimation highlighting how these methods handle nuisance parameters in Sect.~\ref{sec:methods}. Our results and conclusions  can be found in Sect.~\ref{sec:results} and Sect.~\ref{sec:conclusions} respectively.

\section{Data and Model for Spectral time lags}
\label{sec:data}
The observed spectral time lag from a given GRB can be written down as:
\begin{equation}
\Delta t_{obs} = \Delta t_{int} + \Delta t_{LIV},
\label{eq:sum}
\end{equation}
where \(\Delta t_{int}\) is the intrinsic time lag between the emission of photon of a particular energy and the lowest energy photon from the GRB and $\Delta t_{LIV}$ is the LIV-induced time-lag.  W17  used the following model for  the intrinsic emission delay:
\begin{equation}
\Delta t_{int} (E) \rm{(sec)}  = \tau \left[\left(\frac{E}{keV}\right)^\alpha -  \left(\frac{E_0}{keV}\right)^\alpha \right] , 
\label{eq:int}
\end{equation}
\noindent where $E_0$=11.34 keV; whereas  $\tau$ and $\alpha$ are free parameters. This model has been subsequently used  to model the intrinsic lag in many other searches for LIV~\cite{Desairev}.
The  remaining LIV-induced time lag in Eq.~\ref{eq:sum} ($\Delta t_{LIV}$) can be written for sub-luminal LIV  as follows~\cite{Jacob}:
\begin{equation}
\Delta t_{LIV} = -\frac{1+n}{2H_0} \dfrac{E^n -E_0^n}{E^n_{QG,n}} \int_{0}^{z} \dfrac{(1+z^\prime)^n dz^\prime}{\sqrt{\Omega_M (1+z^\prime)^3 + 1-\Omega_M}},
\label{eq:LIV}
\end{equation}
where $E_{QG,n}$ is the Lorentz-violating or quantum gravity scale, above which Lorentz violation kicks in; $H_0$ is the Hubble constant and $\Omega_M$ is the cosmological matter density. W17 considered two different models for LIV, which have $n=1$ and $n=2$, corresponding to linear and quadratic LIV, respectively.  For the cosmological parameters in Eq.~\ref{eq:LIV}, W17 used  $H_0 = 67.3$ km/sec/Mpc and $\Omega_M$=0.315.

W17 considered the spectra lags for GRB 160625B, located at redshift of $z=1.41$. W17 collated 37 spectral lags using data from Fermi-GBM and Fermi-LAT relative to the  lowest energy band of 10-12 keV, extending up to 20 MeV (cf. Table 1 of W17). W17 then used Bayesian inference, where the sampling of the posterior was done using Markov Chain Monte Carlo (MCMC). W17 obtained  lower limit on $E_{QG}$ given by  $E_{QG} \geq 0.5 \times 10^{16}$ GeV and $E_{QG} \geq 1.4 \times 10^7$ GeV for linear and quadratic LIV, respectively, at $1\sigma$. Bayesian parameter estimation for the same dataset has also been done using  Variational Inference~\cite{Gunapati}. Using both these methods, closed $1\sigma$ bounded intervals for $E_{QG}$ were obtained, implying that prima-facie a central interval should be quoted for $E_{QG}$ instead of a one-sided lower limit.

\section{Comparison of Bayesian and frequentist inference}
\label{sec:methods}
We now provide a very  brief primer on Bayesian and frequentist parameter estimation and highlight some of the differences between the two methods for our particular use case. More details on Bayesian parameter estimation can be found in recent reviews~\cite{Trotta,Sanjib,Krishak}. Frequentist parameter estimation is usually reviewed in Particle Data Group, with the latest update in ~\cite{PDG}. 

For both these methods, one needs to model the probability of the data  ($D$) given a parametric function consisting of parameter vector ($\theta$).
We denote this probability by $P(D|\theta)$. For our example, this has been modeled by a Gaussian likelihood ($\mathcal{L}(\theta) $) as follows:
\begin{equation}
     P(D|\theta)=\mathcal{L}(\theta) =\prod_{i=1}^N \frac{1}{\sigma_{i} \sqrt{2\pi}} \exp \left\{-\frac{[\Delta t _i-f(\Delta E_i,\theta)]^2}{2\sigma_{i}^2}\right\},
     \label{eq:likelihood}
  \end{equation}
where $N$ is the total number of data points; $\Delta t_i$ denotes the data which correspond to the observed  spectral lags, and  $\sigma_{i}$ denotes the observed uncertainty in the spectral lag. The function $f(\Delta E_i,\theta)$ is obtained from the sum of Eq.~\ref{eq:int} and Eq.~\ref{eq:LIV}. 

In Bayesian inference, one evaluates the Bayesian Posterior $P(\theta|D)$ which is given by $P(\theta|D) \propto P(D|\theta) P(\theta)$, where $P(\theta)$ is the prior on parameter vector $\theta$. Bayesian parameter inference then entails obtaining central estimates from the posterior probability distribution. In practice, almost all Bayesian computations are nowadays done using MCMC (although see ~\cite{Gunapati}), and the median estimator along with the 68 percentile intervals are computed from the MCMC chains to obtain marginalized 1$\sigma$ intervals~\cite{Sanjib}.  

Usually the parameter vector $\theta$ consists of more than one free  parameter.  Among these, we might be most interested in only one of the parameters. In such cases, the other free parameters can be  considered as nuisance parameters. For our particular use case, $\theta$
consists of  three parameters: \{$E_{QG}$,$\tau$,$\alpha$\}. Since we are mainly interested in constraining $E_{QG}$, the astrophysical parameters $\tau$ and $\alpha$ can be considered as nuisance parameters. For the sake of illustration, let us assume that  in a generic setting the parameter vector ($\theta$) consists of two parameters: $\theta$ =\{$\phi$,$\alpha$\}. Among these, let us consider $\phi$ to be the parameter of interest and $\alpha$ to be the nuisance parameter. In Bayesian inference, the central estimates for $\phi$ are obtained by integrating the posterior over the nuisance parameter $\alpha$ to get the posterior distribution for $P(\phi)$.
\begin{equation}
P(\phi)= \int P(\phi,\alpha|D) d\alpha,
\end{equation}
where $P(\phi,\alpha|D)$ is the posterior for $\theta$. This process is known as marginalization. The central estimates and error intervals are obtained from $P(\phi)$. All previous works on searches for LIV along with the constraints on $E_{QG}$ have always followed the above prescription~\cite{Desairev}.

To deal with nuisance parameters in frequentist statistics on the other hand, one calculates the profile likelihood, obtained  by maximizing the combined likelihood $\mathcal{L}(\phi,\alpha)$ with respect to $\alpha$:
\begin{equation}
\mathcal{L}(\phi)= \max_{\alpha} \mathcal{L}(\phi,\alpha)
\label{eq:pL}
\end{equation}
The central estimate for $\phi$ can then be obtained from $\mathcal{L}(\phi)$. In practice, $\chi^2(\phi) \equiv -2\ln  \mathcal{L}(\phi)$ is defined, and frequentist confidence intervals are constructed from $\Delta \chi^2 (\phi) = \chi^2 (\phi) - \chi^2_{min}$, where $\chi^2_{min}$ is the global minimum for $\chi^2(\phi)$. \rthis{According to Wilks' theorem, $\Delta\chi^2$ follows a $\chi^2$ distribution  for one degree of freedom~\cite{Wilks1938,Herold24}.}
If  $\chi^2_{min}$  is far from the physical boundary, the central interval for the parameter $\phi$ at a given confidence level can be obtained using Newman prescription from the $\Delta \chi^2$ intercept~\cite{NR}. Close to the physical boundary one must use the Feldman-Cousins prescription~\cite{FC}.

This method of profile likelihood has many potential differences compared to the Bayesian counterpart~\cite{Cousins95}. The profile likelihood does not require priors unlike Bayesian inference, which could affect the final results. The profile likelihood formalism also allows  us to include the effect of physical boundary using the Feldman-Cousins prescription~\cite{FC}. The profile likelihood also  does not suffer from the volume effect, which could arise in marginalization~\cite{Gomez}.  Other advantages of profile likelihood over Bayesian analyses have been extensively discussed in recent works related to parameter estimation in Cosmology~\cite{Campeti,Herold24,Gomez}. Most recently, this concept of profiling over nuisance parameters has also been applied to the Bayesian posterior to define a ``profile posterior''~\cite{Weller,Raveri}. This is a hybrid method combining the tenets of both frequentist and Bayesian analysis.

We now apply the profile likelihood  method to the spectral lag data for GRB 1606025B in order to constrain $E_{QG}$.

\section{Application of Profile likelihood to  GRB 1606025B spectral lag data}
\label{sec:results}
\begin{figure}
\centering
\includegraphics[width=0.5\textwidth]{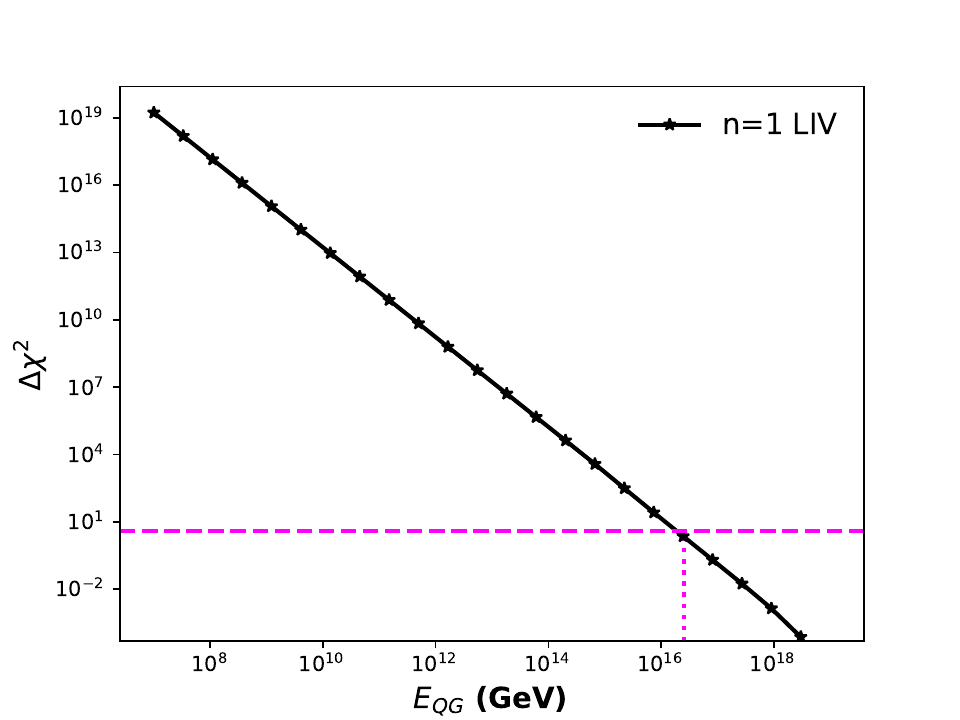}
\caption{$\Delta \chi^2$ (defined as $\chi^2-\chi^2_{min}$) as a function of $E_{QG}$ for linear LIV, corresponding to $n=1$ in Eq.~\ref{eq:LIV}. The horizontal  magenta dashed line is at $\Delta \chi^2=4$ and the corresponding X-intercept of the curves (magenta dotted line) gives the \rthis{95\% c.l.} lower limit at $E_{QG}=2.55 \times 10^{16}$ GeV. }
\label{fig1}
\end{figure}

\begin{figure}
\centering
\includegraphics[width=0.5\textwidth]{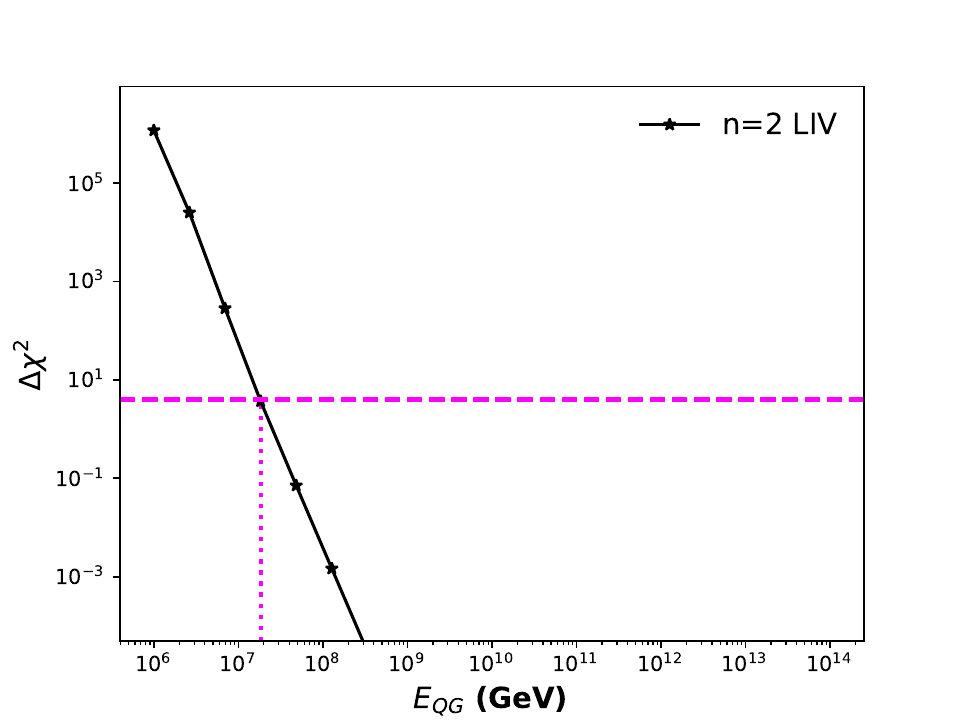}
\caption{$\Delta \chi^2$ (defined as $\chi^2-\chi^2_{min}$)  as a function of $E_{QG}$ for quadratic LIV, corresponding to $n=2$ in Eq.~\ref{eq:LIV}. The horizontal magenta dashed line is at $\Delta \chi^2=4.0$ and the corresponding X-intercept of the curves (magenta dotted line)  gives the \rthis{95\% c.l.} lower limit at $E_{QG}=1.85 \times 10^{7}$ GeV.  }
\label{fig2}
\end{figure}

For both linear and quadratic LIV, our parameter vector consists of three parameters \{$E_{QG}$,$\tau$,$\alpha$\}, where we  are mostly interested in the estimates of $E_{QG}$. Therefore, $\tau$ and $\alpha$ can be considered as nuisance parameters. We use the same likelihood as in Eq.~\ref{eq:likelihood}. To simplify the calculation of  the profile likelihood using Eq.~\ref{eq:pL}, we minimize $\chi^2$ obtained the full likelihood, given by $\chi2\equiv -2 \ln \mathcal{L}(\theta)$, using the full likelihood defined in Eq.~\ref{eq:likelihood}.
We then construct  a logarithmically spaced grid for $E_{QG}$ from $10^6$ to $10^{19}$ GeV.  The upper bound of $10^{19}$ GeV corresponds to the  Planck scale and can be considered as the physical boundary.  For each value of $E_{QG}$ at this grid, we calculate the minimum value of $\chi^2 (E_{QG})$ by minimizing over $\alpha$ and $\tau$. For this purpose we used {\tt scipy.optimize.fmin} function, which uses the Nelder-Mead simplex algorithm~\cite{NR}. As a cross-check we also compared with the Powell minimization algorithm built in {\tt scipy}, which gives the same results. Therefore, the global minimum for $\chi^2$ is robust. We then plot $\Delta \chi^2$ as a function of $E_{QG}$, where  
$\Delta \chi^2= \chi^2 (E_{QG}) -\chi^2_{min}$. The corresponding $\Delta \chi^2$ curves as a function of $E_{QG}$ can be found in Fig.~\ref{fig1} and   Fig.~\ref{fig2} for linear and quadratic LIV, respectively. For both the LIV models we find that $\Delta \chi^2$ always decreases with increasing $E_{QG}$. Here, $\chi^2_{min}$ corresponds to $\chi^2$ at the Planck scale, which can be considered as the physical boundary.  Therefore, there is no  global minimum for $\Delta \chi^2$ followed by a rising trend. This is different from previous results obtained using Bayesian inference, where  closed $1\sigma$ intervals for  $E_{QG}$ were obtained after  marginalizing over $\tau$ and $\alpha$~\cite{Wei,Gunapati}.

Therefore, we can obtain a one-sided lower limit on $E_{QG}$ in a seamless way.  Based on the Newman prescription the \rthis{95.4\%  (95\%,} to shorten the notation) lower limit is given by the value of $E_{QG}$ for which  $\Delta \chi^2=4.0$~\cite{NR,Herold}. Therefore, the \rthis{95\%} lower limits  $E_{QG} \ge 2.55 \times 10^{16}$ GeV and $E_{QG} \ge 1.85 \times 10^{7}$ GeV for linear and quadratic LIV, respectively.   We note that since the $E_{QG}$ values for  $\Delta \chi^2=4$   are obtained far from the physical boundary, the Newman prescription suffices and there  is  no need to switch to the Feldman-Cousins prescription. 

\section{Conclusions}
\label{sec:conclusions}
In this work we have reanalyzed the data for spectral lag transition in GRB 1606025B, using the  frequentist method  of profile likelihood in order to constrain the energy scale for LIV. All previous searches for LIV using GRB spectral lags have used Bayesian inference, which involved  marginalizing over the astrophysical nuisance parameters~\cite{Wei,Gunapati}. Similar to previous works, we model the spectral lags as a sum of astrophysical induced time lag and LIV induced time lag. We consider the same parametric models for both the lags as in previous works~\cite{Wei,Ganguly,Gunapati}. The astrophysical induced lag (cf. Eq.~\ref{eq:int}) consists of two nuisance parameters, whereas the physically interesting parameter we want to constrain is the energy scale of LIV (denoted by $E_{QG}$ in  Eq.~\ref{eq:LIV}).

For both the LIV models, we calculated the $\Delta \chi^2$ as a function of $E_{QG}$ by computing the minimum value of $\chi^2$ over the astrophysical parameters for each value of $E_{QG}$. These plots of $\Delta \chi^2$ can be found in Fig~\ref{fig1} and ~\ref{fig2} for linear and quadratic LIV, respectively. One difference compared to Bayesian inference is that we do not get a convex shape for the probability distribution for $E_{QG}$ below the Planck scale using the profile likelihood method. Therefore, there is no global minimum   and one can unhesitatingly set one-sided lower limits  on $E_{QG}$ for a given confidence level. The corresponding 95\% lower limits on $E_{QG}$ which we obtain are given by $E_{QG} \ge 2.55 \times 10^{16}$ GeV and $E_{QG} \ge 1.85 \times 10^{7}$ GeV for linear and quadratic LIV, respectively. 
\rthis{In the spirit of open science, we have made our analysis code and data publicly available, which can be found at \url{https://github.com/shantanu9847/LIVPL}.}

Therefore, this is the first proof of principles application of profile likelihood in the analysis of GRB spectral lag data  to search for LIV and provides a seamless way to set a lower limit. In future works,  we shall apply this method to other searches for LIV using GRB spectral lags.  
\begin{acknowledgements}
This work was motivated following very interesting seminars at IIT Hyderabad by Eoin Colgain and Laura Herold. We are grateful to both of them as well as Bob Cousins  for useful discussions.
\end{acknowledgements}

\bibliography{main}

\begin{thebibliography}{27}
\expandafter\ifx\csname natexlab\endcsname\relax\def\natexlab#1{#1}\fi
\expandafter\ifx\csname bibnamefont\endcsname\relax
  \def\bibnamefont#1{#1}\fi
\expandafter\ifx\csname bibfnamefont\endcsname\relax
  \def\bibfnamefont#1{#1}\fi
\expandafter\ifx\csname citenamefont\endcsname\relax
  \def\citenamefont#1{#1}\fi
\expandafter\ifx\csname url\endcsname\relax
  \def\url#1{\texttt{#1}}\fi
\expandafter\ifx\csname urlprefix\endcsname\relax\def\urlprefix{URL }\fi
\providecommand{\bibinfo}[2]{#2}
\providecommand{\eprint}[2][]{\url{#2}}

\bibitem[{\citenamefont{{Desai}}(2024)}]{Desairev}
\bibinfo{author}{\bibfnamefont{S.}~\bibnamefont{{Desai}}}, in \emph{\bibinfo{booktitle}{Recent Progress on Gravity Tests. Challenges and Future Perspectives}}, edited by \bibinfo{editor}{\bibfnamefont{C.}~\bibnamefont{{Bambi}}} \bibnamefont{and} \bibinfo{editor}{\bibfnamefont{A.}~\bibnamefont{{C{\'a}rdenas-Avenda{\~n}o}}} (\bibinfo{year}{2024}), pp. \bibinfo{pages}{433--463}.

\bibitem[{\citenamefont{{Yu} et~al.}(2022)\citenamefont{{Yu}, {Gao}, {Wang}, and {Zhang}}}]{WuGRBreview}
\bibinfo{author}{\bibfnamefont{Y.-W.} \bibnamefont{{Yu}}}, \bibinfo{author}{\bibfnamefont{H.}~\bibnamefont{{Gao}}}, \bibinfo{author}{\bibfnamefont{F.-Y.} \bibnamefont{{Wang}}}, \bibnamefont{and} \bibinfo{author}{\bibfnamefont{B.-B.} \bibnamefont{{Zhang}}}, in \emph{\bibinfo{booktitle}{Handbook of X-ray and Gamma-ray Astrophysics. Edited by Cosimo Bambi and Andrea Santangelo}} (\bibinfo{year}{2022}), p.~\bibinfo{pages}{31}.

\bibitem[{\citenamefont{{Wei} and {Wu}}(2022)}]{WeiWu2}
\bibinfo{author}{\bibfnamefont{J.-J.} \bibnamefont{{Wei}}} \bibnamefont{and} \bibinfo{author}{\bibfnamefont{X.-F.} \bibnamefont{{Wu}}}, in \emph{\bibinfo{booktitle}{Handbook of X-ray and Gamma-ray Astrophysics. Edited by Cosimo Bambi and Andrea Santangelo}} (\bibinfo{year}{2022}), p.~\bibinfo{pages}{82}.

\bibitem[{\citenamefont{{Amelino-Camelia} et~al.}(1998)\citenamefont{{Amelino-Camelia}, {Ellis}, {Mavromatos}, {Nanopoulos}, and {Sarkar}}}]{AmelinoCamelia98}
\bibinfo{author}{\bibfnamefont{G.}~\bibnamefont{{Amelino-Camelia}}}, \bibinfo{author}{\bibfnamefont{J.}~\bibnamefont{{Ellis}}}, \bibinfo{author}{\bibfnamefont{N.~E.} \bibnamefont{{Mavromatos}}}, \bibinfo{author}{\bibfnamefont{D.~V.} \bibnamefont{{Nanopoulos}}}, \bibnamefont{and} \bibinfo{author}{\bibfnamefont{S.}~\bibnamefont{{Sarkar}}}, \bibinfo{journal}{\nat} \textbf{\bibinfo{volume}{393}}, \bibinfo{pages}{763} (\bibinfo{year}{1998}), \eprint{astro-ph/9712103}.

\bibitem[{\citenamefont{{Wei} et~al.}(2017)\citenamefont{{Wei}, {Zhang}, {Shao}, {Wu}, and {M{\'e}sz{\'a}ros}}}]{Wei}
\bibinfo{author}{\bibfnamefont{J.-J.} \bibnamefont{{Wei}}}, \bibinfo{author}{\bibfnamefont{B.-B.} \bibnamefont{{Zhang}}}, \bibinfo{author}{\bibfnamefont{L.}~\bibnamefont{{Shao}}}, \bibinfo{author}{\bibfnamefont{X.-F.} \bibnamefont{{Wu}}}, \bibnamefont{and} \bibinfo{author}{\bibfnamefont{P.}~\bibnamefont{{M{\'e}sz{\'a}ros}}}, \bibinfo{journal}{\apjl} \textbf{\bibinfo{volume}{834}}, \bibinfo{eid}{L13} (\bibinfo{year}{2017}), \eprint{1612.09425}.

\bibitem[{\citenamefont{{Ganguly} and {Desai}}(2017)}]{Ganguly}
\bibinfo{author}{\bibfnamefont{S.}~\bibnamefont{{Ganguly}}} \bibnamefont{and} \bibinfo{author}{\bibfnamefont{S.}~\bibnamefont{{Desai}}}, \bibinfo{journal}{Astroparticle Physics} \textbf{\bibinfo{volume}{94}}, \bibinfo{pages}{17} (\bibinfo{year}{2017}), \eprint{1706.01202}.

\bibitem[{\citenamefont{{Gunapati} et~al.}(2022)\citenamefont{{Gunapati}, {Jain}, {Srijith}, and {Desai}}}]{Gunapati}
\bibinfo{author}{\bibfnamefont{G.}~\bibnamefont{{Gunapati}}}, \bibinfo{author}{\bibfnamefont{A.}~\bibnamefont{{Jain}}}, \bibinfo{author}{\bibfnamefont{P.~K.} \bibnamefont{{Srijith}}}, \bibnamefont{and} \bibinfo{author}{\bibfnamefont{S.}~\bibnamefont{{Desai}}}, \bibinfo{journal}{\pasa} \textbf{\bibinfo{volume}{39}}, \bibinfo{eid}{e001} (\bibinfo{year}{2022}), \eprint{1803.06473}.

\bibitem[{\citenamefont{{Agrawal} et~al.}(2021)\citenamefont{{Agrawal}, {Singirikonda}, and {Desai}}}]{Agrawal_2021}
\bibinfo{author}{\bibfnamefont{R.}~\bibnamefont{{Agrawal}}}, \bibinfo{author}{\bibfnamefont{H.}~\bibnamefont{{Singirikonda}}}, \bibnamefont{and} \bibinfo{author}{\bibfnamefont{S.}~\bibnamefont{{Desai}}}, \bibinfo{journal}{\jcap} \textbf{\bibinfo{volume}{2021}}, \bibinfo{eid}{029} (\bibinfo{year}{2021}), \eprint{2102.11248}.

\bibitem[{\citenamefont{{Desai} et~al.}(2023)\citenamefont{{Desai}, {Agrawal}, and {Singirikonda}}}]{Desai23}
\bibinfo{author}{\bibfnamefont{S.}~\bibnamefont{{Desai}}}, \bibinfo{author}{\bibfnamefont{R.}~\bibnamefont{{Agrawal}}}, \bibnamefont{and} \bibinfo{author}{\bibfnamefont{H.}~\bibnamefont{{Singirikonda}}}, \bibinfo{journal}{European Physical Journal C} \textbf{\bibinfo{volume}{83}}, \bibinfo{eid}{63} (\bibinfo{year}{2023}), \eprint{2205.12780}.

\bibitem[{\citenamefont{{Pasumarti} and {Desai}}(2023)}]{Pasumarti23}
\bibinfo{author}{\bibfnamefont{V.}~\bibnamefont{{Pasumarti}}} \bibnamefont{and} \bibinfo{author}{\bibfnamefont{S.}~\bibnamefont{{Desai}}}, \bibinfo{journal}{Journal of High Energy Astrophysics} \textbf{\bibinfo{volume}{40}}, \bibinfo{pages}{41} (\bibinfo{year}{2023}), \eprint{2307.02296}.

\bibitem[{\citenamefont{{Particle Data Group} et~al.}(2020)\citenamefont{{Particle Data Group}, {Zyla}, {Barnett}, {Beringer}, {Dahl}, {Dwyer}, {Groom}, {Lin}, {Lugovsky}, {Pianori} et~al.}}]{PDG}
\bibinfo{author}{\bibnamefont{{Particle Data Group}}}, \bibinfo{author}{\bibfnamefont{P.~A.} \bibnamefont{{Zyla}}}, \bibinfo{author}{\bibfnamefont{R.~M.} \bibnamefont{{Barnett}}}, \bibinfo{author}{\bibfnamefont{J.}~\bibnamefont{{Beringer}}}, \bibinfo{author}{\bibfnamefont{O.}~\bibnamefont{{Dahl}}}, \bibinfo{author}{\bibfnamefont{D.~A.} \bibnamefont{{Dwyer}}}, \bibinfo{author}{\bibfnamefont{D.~E.} \bibnamefont{{Groom}}}, \bibinfo{author}{\bibfnamefont{C.~J.} \bibnamefont{{Lin}}}, \bibinfo{author}{\bibfnamefont{K.~S.} \bibnamefont{{Lugovsky}}}, \bibinfo{author}{\bibfnamefont{E.}~\bibnamefont{{Pianori}}}, \bibnamefont{et~al.}, \bibinfo{journal}{Progress of Theoretical and Experimental Physics} \textbf{\bibinfo{volume}{2020}}, \bibinfo{eid}{083C01} (\bibinfo{year}{2020}).

\bibitem[{\citenamefont{{Herold} et~al.}(2022)\citenamefont{{Herold}, {Ferreira}, and {Komatsu}}}]{Herold}
\bibinfo{author}{\bibfnamefont{L.}~\bibnamefont{{Herold}}}, \bibinfo{author}{\bibfnamefont{E.~G.~M.} \bibnamefont{{Ferreira}}}, \bibnamefont{and} \bibinfo{author}{\bibfnamefont{E.}~\bibnamefont{{Komatsu}}}, \bibinfo{journal}{\apjl} \textbf{\bibinfo{volume}{929}}, \bibinfo{eid}{L16} (\bibinfo{year}{2022}), \eprint{2112.12140}.

\bibitem[{\citenamefont{{Campeti} and {Komatsu}}(2022)}]{Campeti}
\bibinfo{author}{\bibfnamefont{P.}~\bibnamefont{{Campeti}}} \bibnamefont{and} \bibinfo{author}{\bibfnamefont{E.}~\bibnamefont{{Komatsu}}}, \bibinfo{journal}{\apj} \textbf{\bibinfo{volume}{941}}, \bibinfo{eid}{110} (\bibinfo{year}{2022}), \eprint{2205.05617}.

\bibitem[{\citenamefont{{Colg{\'a}in} et~al.}(2024)\citenamefont{{Colg{\'a}in}, {Pourojaghi}, and {Sheikh-Jabbari}}}]{Colgain24}
\bibinfo{author}{\bibfnamefont{E.~{\'O}.} \bibnamefont{{Colg{\'a}in}}}, \bibinfo{author}{\bibfnamefont{S.}~\bibnamefont{{Pourojaghi}}}, \bibnamefont{and} \bibinfo{author}{\bibfnamefont{M.~M.} \bibnamefont{{Sheikh-Jabbari}}}, \bibinfo{journal}{arXiv e-prints} \bibinfo{eid}{arXiv:2406.06389} (\bibinfo{year}{2024}), \eprint{2406.06389}.

\bibitem[{\citenamefont{{Karwal} et~al.}(2024)\citenamefont{{Karwal}, {Patel}, {Bartlett}, {Poulin}, {Smith}, and {Pfeffer}}}]{Karwal24}
\bibinfo{author}{\bibfnamefont{T.}~\bibnamefont{{Karwal}}}, \bibinfo{author}{\bibfnamefont{Y.}~\bibnamefont{{Patel}}}, \bibinfo{author}{\bibfnamefont{A.}~\bibnamefont{{Bartlett}}}, \bibinfo{author}{\bibfnamefont{V.}~\bibnamefont{{Poulin}}}, \bibinfo{author}{\bibfnamefont{T.~L.} \bibnamefont{{Smith}}}, \bibnamefont{and} \bibinfo{author}{\bibfnamefont{D.~N.} \bibnamefont{{Pfeffer}}}, \bibinfo{journal}{arXiv e-prints} \bibinfo{eid}{arXiv:2401.14225} (\bibinfo{year}{2024}), \eprint{2401.14225}.

\bibitem[{\citenamefont{{Herold} et~al.}(2024)\citenamefont{{Herold}, {Ferreira}, and {Heinrich}}}]{Herold24}
\bibinfo{author}{\bibfnamefont{L.}~\bibnamefont{{Herold}}}, \bibinfo{author}{\bibfnamefont{E.~G.~M.} \bibnamefont{{Ferreira}}}, \bibnamefont{and} \bibinfo{author}{\bibfnamefont{L.}~\bibnamefont{{Heinrich}}}, \bibinfo{journal}{arXiv e-prints} \bibinfo{eid}{arXiv:2408.07700} (\bibinfo{year}{2024}), \eprint{2408.07700}.

\bibitem[{\citenamefont{{Jacob} and {Piran}}(2008)}]{Jacob}
\bibinfo{author}{\bibfnamefont{U.}~\bibnamefont{{Jacob}}} \bibnamefont{and} \bibinfo{author}{\bibfnamefont{T.}~\bibnamefont{{Piran}}}, \bibinfo{journal}{\jcap} \textbf{\bibinfo{volume}{1}}, \bibinfo{eid}{031} (\bibinfo{year}{2008}), \eprint{0712.2170}.

\bibitem[{\citenamefont{{Trotta}}(2017)}]{Trotta}
\bibinfo{author}{\bibfnamefont{R.}~\bibnamefont{{Trotta}}}, \bibinfo{journal}{ArXiv e-prints}  (\bibinfo{year}{2017}), \eprint{1701.01467}.

\bibitem[{\citenamefont{{Sharma}}(2017)}]{Sanjib}
\bibinfo{author}{\bibfnamefont{S.}~\bibnamefont{{Sharma}}}, \bibinfo{journal}{\araa} \textbf{\bibinfo{volume}{55}}, \bibinfo{pages}{213} (\bibinfo{year}{2017}), \eprint{1706.01629}.

\bibitem[{\citenamefont{{Krishak} and {Desai}}(2020)}]{Krishak}
\bibinfo{author}{\bibfnamefont{A.}~\bibnamefont{{Krishak}}} \bibnamefont{and} \bibinfo{author}{\bibfnamefont{S.}~\bibnamefont{{Desai}}}, \bibinfo{journal}{\jcap} \textbf{\bibinfo{volume}{2020}}, \bibinfo{eid}{006} (\bibinfo{year}{2020}), \eprint{2003.10127}.

\bibitem[{\citenamefont{Wilks}(1938)}]{Wilks1938}
\bibinfo{author}{\bibfnamefont{S.~S.} \bibnamefont{Wilks}}, \bibinfo{journal}{The annals of mathematical statistics} \textbf{\bibinfo{volume}{9}}, \bibinfo{pages}{60} (\bibinfo{year}{1938}).

\bibitem[{\citenamefont{{Press} et~al.}(1992)\citenamefont{{Press}, {Teukolsky}, {Vetterling}, and {Flannery}}}]{NR}
\bibinfo{author}{\bibfnamefont{W.~H.} \bibnamefont{{Press}}}, \bibinfo{author}{\bibfnamefont{S.~A.} \bibnamefont{{Teukolsky}}}, \bibinfo{author}{\bibfnamefont{W.~T.} \bibnamefont{{Vetterling}}}, \bibnamefont{and} \bibinfo{author}{\bibfnamefont{B.~P.} \bibnamefont{{Flannery}}}, \emph{\bibinfo{title}{{Numerical recipes in FORTRAN. The art of scientific computing}}} (\bibinfo{year}{1992}).

\bibitem[{\citenamefont{{Feldman} and {Cousins}}(1998)}]{FC}
\bibinfo{author}{\bibfnamefont{G.~J.} \bibnamefont{{Feldman}}} \bibnamefont{and} \bibinfo{author}{\bibfnamefont{R.~D.} \bibnamefont{{Cousins}}}, \bibinfo{journal}{\prd} \textbf{\bibinfo{volume}{57}}, \bibinfo{pages}{3873} (\bibinfo{year}{1998}), \eprint{physics/9711021}.

\bibitem[{\citenamefont{{Cousins}}(1995)}]{Cousins95}
\bibinfo{author}{\bibfnamefont{R.~D.} \bibnamefont{{Cousins}}}, \bibinfo{journal}{American Journal of Physics} \textbf{\bibinfo{volume}{63}}, \bibinfo{pages}{398} (\bibinfo{year}{1995}).

\bibitem[{\citenamefont{{G{\'o}mez-Valent}}(2022)}]{Gomez}
\bibinfo{author}{\bibfnamefont{A.}~\bibnamefont{{G{\'o}mez-Valent}}}, \bibinfo{journal}{\prd} \textbf{\bibinfo{volume}{106}}, \bibinfo{eid}{063506} (\bibinfo{year}{2022}), \eprint{2203.16285}.

\bibitem[{\citenamefont{{Kerscher} and {Weller}}(2024)}]{Weller}
\bibinfo{author}{\bibfnamefont{M.}~\bibnamefont{{Kerscher}}} \bibnamefont{and} \bibinfo{author}{\bibfnamefont{J.}~\bibnamefont{{Weller}}}, \bibinfo{journal}{arXiv e-prints} \bibinfo{eid}{arXiv:2408.02063} (\bibinfo{year}{2024}), \eprint{2408.02063}.

\bibitem[{\citenamefont{{Raveri} et~al.}(2024)\citenamefont{{Raveri}, {Doux}, and {Pandey}}}]{Raveri}
\bibinfo{author}{\bibfnamefont{M.}~\bibnamefont{{Raveri}}}, \bibinfo{author}{\bibfnamefont{C.}~\bibnamefont{{Doux}}}, \bibnamefont{and} \bibinfo{author}{\bibfnamefont{S.}~\bibnamefont{{Pandey}}}, \bibinfo{journal}{arXiv e-prints} \bibinfo{eid}{arXiv:2409.09101} (\bibinfo{year}{2024}), \eprint{2409.09101}.

\end{thebibliography}
\end{document}